%% file: man.tex
\documentclass[aps,prl,twocolumn,groupedaddress,superscriptaddress]{revtex4}

\usepackage{graphicx,color}
\usepackage[colorlinks=true,urlcolor=blue,citecolor=blue,linkcolor=blue]{hyperref}

\usepackage{amsmath,amssymb, booktabs, braket}

\begin{document}

\title{Measurement-Free Ultrafast Quantum Error Correction by Using Multi-Controlled Gates in Higher-Dimensional State Space}

\author{Toshiaki~Inada$^1$, Wonho~Jang$^2$, Yutaro~Iiyama$^1$, Koji~Terashi$^1$,\\ Ryu~Sawada$^1$, Junichi~Tanaka$^1$, Shoji~Asai$^{1,2}$}
\affiliation{$^1$International Center for Elementary Particle Physics, The University of Tokyo, 7-3-1 Hongo, Bunkyo-ku, Tokyo 113-0033, Japan}
\affiliation{$^2$Department of Physics, Graduate School of Science, The University of Tokyo, 7-3-1 Hongo, Bunkyo-ku, Tokyo 113-0033, Japan}

\date{\today}

\begin{abstract}
Quantum error correction is a crucial step beyond the current noisy-intermediate-scale quantum device towards fault-tolerant quantum computing.
However, most of the error corrections ever demonstrated rely on post-selection of events or post-correction of states, based on measurement results repeatedly recorded during circuit execution.
On the other hand, real-time error correction is supposed to be performed through classical feedforward of the measurement results to data qubits.
It provides unavoidable latency from conditional electronics that would limit the scalability of the next-generation quantum processors.
Here we propose a new approach to real-time error correction that is free from measurement and realized by using multi-controlled gates based on higher-dimensional state space.
Specifically, we provide a series of novel decompositions of a Toffoli gate by using the lowest three energy levels of a transmon that significantly reduce the number of two-qubit gates and discuss their essential features, such as extendability to an arbitrary number of control qubits, the necessity of exclusively controlled NOT gates, and usefulness of their incomplete variants.
Combined with the recently demonstrated schemes of fast two-qubit gates and all-microwave qubit reset, it would substantially shorten the time required for error correction and resetting ancilla qubits compared to a measurement-based approach and provide an error correction rate of $\gtrsim1$~MHz with high accuracy for three-qubit bit- and phase-flip errors.
\end{abstract}

\maketitle

\begin{figure}[b]
\vspace{+2mm}
\includegraphics[width=85mm]{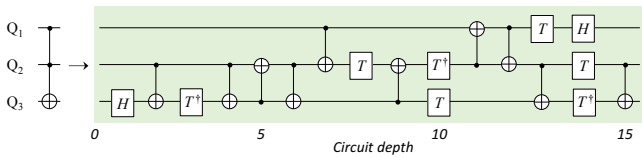}
\caption{
Typical Toffoli gate transpilation using nine single-qubit gates and 10 CNOT gates with a circuit depth of 15.
$H$ and $T$ are a Hadamard and a $\pi /4$ phase gate, respectively.
\label{fig1}}
\end{figure}

\paragraph{Introduction.---}
Multi-controlled gates are important quantum operations in many aspects of advanced quantum information processing, such as quantum error correction (QEC)~\cite{preskill2018quantum,shor1995,calderbank1996good,terhal2015quantum}, fault tolerance~\cite{jones2013ft-toffoli,paetznick2013,yoder2016}, gate universality~\cite{shi2003,aharonov2003}, and quantum algorithms~\cite{shor1999polynomial,grover1997,hner2017}.
Naively realizing them in a similar approach that extends the one used in two-qubit gates requires many-body interactions.
When considering their decomposition, even the simplest, two-controlled NOT (Toffoli) gate uses ten controlled-NOT (CNOT) gates and nine single-qubit gates (Fig.~\ref{fig1}).
Thus, extending the number of control qubits in a multi-controlled gate requires a different class of quantum logic and becomes the bottleneck in many quantum algorithms~\cite{shor1999polynomial,grover1997,hner2017,childs2010,montanaro2016,mcardle2020}.

So far, many schemes have been proposed to realize the functionality of multi-controlled gates.
They need, for example, fine-tuning of device parameters, an elaborate layout of qubits, or optimal control of drive waveforms in a specific coupling configuration~\cite{banchi2016,rasmussen2020,stojanovic2012,zahedinejad2015}.
Without lowering the flexibility in circuit design, here we show an efficient and more straightforward way of realizing multi-controlled gates in higher-dimensional state space inherent to most quantum systems.
By utilizing this degree of freedom as a precious computational resource, we derive a series of novel Toffoli decompositions that can be extended to an arbitrary number of control qubits and discuss their differences in terms of a circuit depth, a gate time during which higher states are occupied, the necessity of pure CNOTs, and the availability of two-qubit gates in higher-dimensional state space in the first part of this work.

\begin{figure*}[t]
\includegraphics[width=150mm]{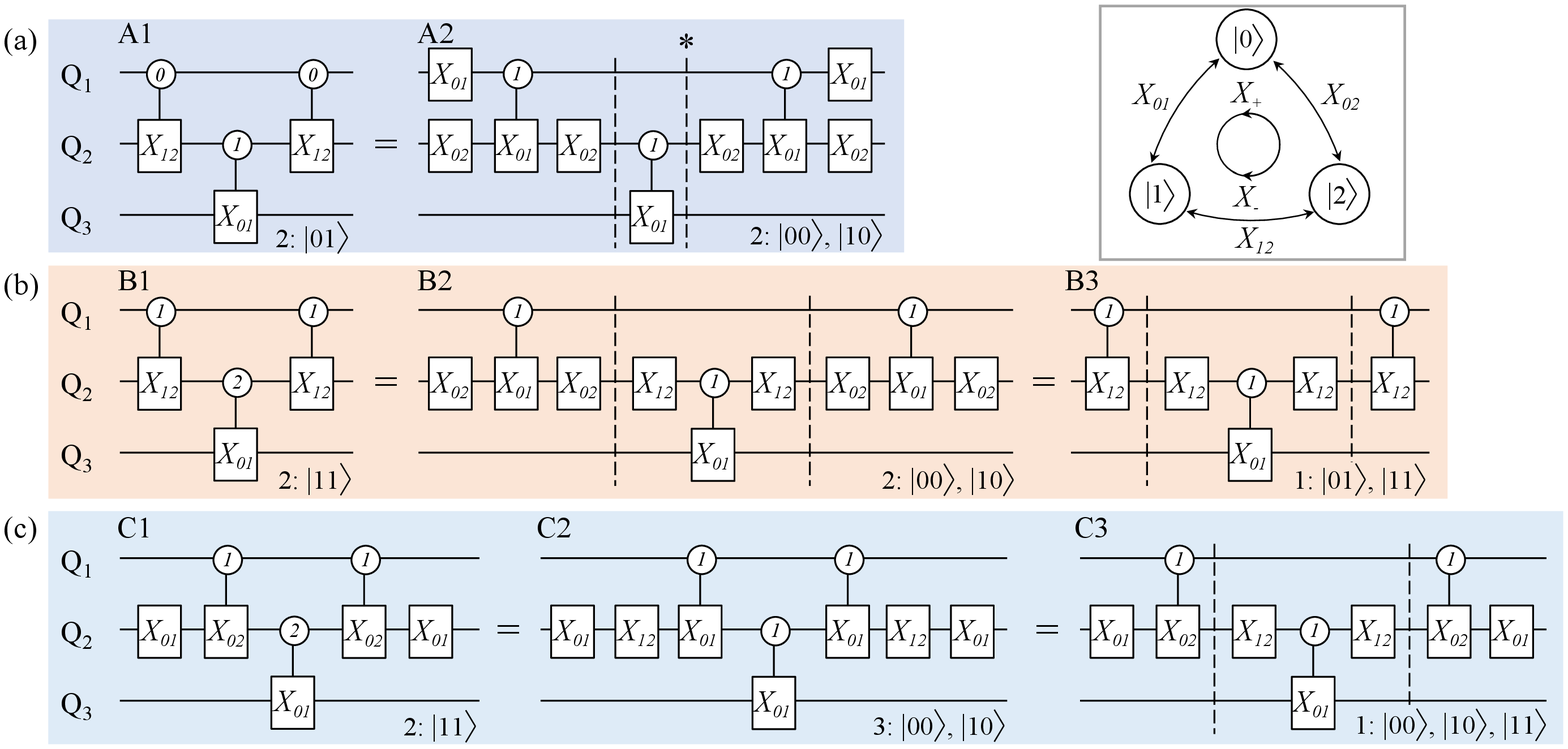}
\caption{
Qutrit-based Toffoli gate decompositions by using three controlled-qutrit gates.
Q$_1$ and Q$_2$ are the two controls, and Q$_3$ is the target.
{\bf (a-c)}~Decompositions into controlled subspace gates CX$_{ij}$ (A1, B1, and C1).
By further decomposing them into the conventional CNOTs ($=\ket{1}$-controlled $X_{01}$ gates), a Toffoli gate can be composed of three CNOTs and several single-subspace gates (A2, B2, and C2).
Partial decompositions are also possible and yield hybrid circuits containing conventional CNOTs and CX$_{ij}$ with shorter circuit depth than the second decompositions (for example, B3 and C3).
For each decomposition, $\tau_{{\rm Q2}=\ket{2},{\rm max}}:\ket{{\rm Q_1Q_2}}$ is shown on the bottom right, where
$\tau_{{\rm Q_2}=\ket{2},{\rm max}}$ is the maximum time among all input states during which ${\rm Q_2}$ stays in $\ket{2}$ for the input state(s) of $\ket{{\rm Q_1Q_2}}$, expressed in the unit of a CNOT gate time by neglecting single-qutrit gate times and considering active CX$_{ij}$ between Q$_1$ and Q$_2$ to be a half.
The vertical lines show the original gate boundaries in the first decompositions as a guide to the eye.
Since the right half of all the decompositions performs uncomputation to restore the input control states, stopping the sequence right after the central gate, for example, at the asterisk ($*$) in A2, leaves Q$_2$ in a qutrit state, giving an {\it incomplete} Toffoli gate with a much shorter overall gate time.
{\bf (Inset)}~Five single-qutrit gates.
Three $X_{ij}$ are the Pauli $X$ gates in the $\ket{i}$-$\ket{j}$ subspace, and $X_{\pm}$ are the cyclic permutations among the qutrit states.
\label{fig2}}
\end{figure*}

In the second part, we show an application of the higher-level-based Toffoli decompositions to QEC.
While QEC experiments have been performed in various quantum systems~\cite{cory1998experimental,knill2001benchmarking,schindler2011experimental,moussa2011demonstration,zhang2011experimental,reed2012realization,zhang2012experimental,bell2014experimental,kelly2015state,nigg2014quantum,waldherr2014quantum,riste2015detecting,corcoles2015demonstration,cramer2016repeated,ofek2016extending,takita2017experimental,linke2017fault,wootton2018repetition,andersen2019entanglement,gong2019experimental,hu2019quantum,wootton2020benchmarking,andersen2020,bultink2020protecting,egan2020fault,luo2020quantum,campagne2020quantum,google2021}
(NMR~\cite{cory1998experimental,knill2001benchmarking,moussa2011demonstration,zhang2011experimental,zhang2012experimental},
ion trap~\cite{schindler2011experimental,nigg2014quantum,linke2017fault,egan2020fault},
photons~\cite{bell2014experimental,luo2020quantum},
NV center~\cite{waldherr2014quantum}, and
3D cavity~\cite{ofek2016extending,hu2019quantum,campagne2020quantum}),
most of the QECs performed on superconducting qubits rely on post-selection of events or post-correction of states, based on measurement results repeatedly recorded during circuit execution~\cite{reed2012realization,kelly2015state,riste2015detecting,corcoles2015demonstration,cramer2016repeated,takita2017experimental,wootton2018repetition,andersen2019entanglement,gong2019experimental,wootton2020benchmarking,andersen2020,bultink2020protecting,google2021}.
On the other hand, real-time error correction is performed through classical feedforward of the measurement results to data qubits and provides unavoidable latency from conditional electronics~\cite{riste2013,salathe2018,andersen2019,bultink2020,riste2020,corcoles2021}, posing a limit in applying the same scheme to the next-generation quantum processors.
Here we propose a new scheme that is free from measurement and realized using the multi-controlled gates derived in the first part.
Combined with an all-microwave reset scheme and fast and high-fidelity two-qubit gates, it can provide real-time QEC with a repetition rate of $\gtrsim1$~MHz for bit- and phase-flip errors in a simple three-qubit protocol.

\paragraph{Qutrit-based Toffoli decomposition.---}
Qudits generally describe the unit of quantum information carried by $d$ orthogonal quantum states~\cite{wang2020}.
The minimal extension of a qubit by adding the second excited state $\ket{2}$ to the computational basis is referred to as qutrit~\cite{di2012}, written as $\ket{\psi} = \alpha \ket{0} + \beta \ket{1} + \gamma \ket{2}$
with $\left| \alpha \right|^2 + \left| \beta \right|^2 + \left| \gamma \right|^2 = 1$.
The addition of $\ket{2}$ significantly increases the number of single qutrit gates to five with three subspace gates ($X_{01}$, $X_{12}$ and $X_{02}$) and two rotational gates ($X_{\pm}$) as in Fig.~\ref{fig2}(inset).
The previous study~\cite{gokhale2019} proposed a Toffoli gate decomposition by using controlled rotational gates, which realization requires further decomposition into either twice the number of controlled qutrit gates or non-trivial single- and two-qutrit gates (see Supplemental Material for details).
Instead, here we propose more efficient decomposition by using three controlled subspace gates CX$_{ij}$ as shown in A1, B1, and C1 of Fig.~\ref{fig2}.
By decomposing them further into the conventional CNOT gate ($=\ket{1}$-controlled $X_{01}$), Toffoli gates can be composed of three CNOTs and several single-subspace gates (A2, B2, and C2).
Note that a two-photon process is required to induce the $X_{02}$ gate~\cite{yurtalan2020}, on which the C2 decomposition does not rely.

Special care needs to be paid when implementing the CX gate in the middle of the sequence with cross-resonance.
If the control Q2 takes the $\ket{2}$ state during this gate, the conventional CNOT gate causes partial Rabi rotation on the target qubit.
This unwanted effect can be removed by tuning the Rabi frequency to the same value as the one conditioned to $\ket{0}$ through their nonlinear dependence on the drive amplitude~\cite{blok2021,tripathi2019,magesan2020}.
Another point on the qutrit operation comes from the enhanced charge dispersion in higher excited states, causing a reduced dephasing time $T_\phi$~\cite{koch2007}.
As the Toffoli decompositions A1, B1, and C1 are composed of two kinds of two-qutrit gates, there are variant decompositions, depending on whether to convert the two-qutrit gate into CNOT, except for A1, which central two-qutrit gate is already CNOT.
B3 and C3 in Fig.~\ref{fig2} show hybrid circuits that minimize the time spent by Q2 in the $\ket{2}$ state, which is expressed in the unit of the CNOT gate time by neglecting single-qutrit ones.

\begin{figure}[t]
\includegraphics[width=87mm]{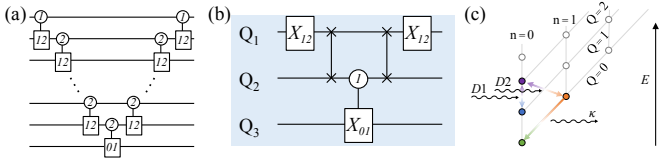}
\caption{
{\bf (a)}~Scalability to the generalized n-controlled Toffoli gate in the case of B1 in Fig.~\ref{fig2}.
A shorthand notation is adopted; for example, 12:~X$_{12}$.
{\bf (b)}~Another Toffoli decomposition by using two $X_{12}$ gates and conventional iSWAPs.
The iSWAPs operate only for $\ket{{\rm Q}_{1}=0,{\rm Q}_{2}=1,{\rm Q}_{3}}$ and give a total phase of $\pi$, which can be compensated by a two-qubit gate, for example, a $\ket{0}$-controlled Z gate.
{\bf (c)}~All-microwave qutrit initialization with the resonator-assisted double drive reset (DDR) scheme~\cite{magnard2018}, where two simultaneous drives $D1$ and $D2$ couple the qutrit excited states $\ket{{\rm Q}=1,n=0}$ and $\ket{{\rm Q}=2,n=0}$ to $\ket{{\rm Q}=0,n=1}$, whose resonator photon decays through a lossy channel with the rate $\kappa$.
\label{fig3}}
\end{figure}

Scalability to the generalized n-controlled Toffoli gate is another essential feature of qutrit-based Toffoli decompositions.
For example, type-B and C decompositions can be linearly extended to an arbitrary number of control qubits as shown in Fig.~\ref{fig3}(a).
Note that controlled subspace gates, for example, $\ket{1}$-controlled $X_{12}$ gates in B1, are possible similarly to the conventional CNOT with cross-resonance, where the control is driven at $\omega_{12}$ instead of $\omega_{01}$.
Determining which decomposition type or the variant is suited for a specific device would depend on the actual nonlinearity of Rabi frequencies in the device, the calibration cost of two-qutrit gates, the overall circuit depth, and the error budget on the Toffoli gate.

\paragraph{iSWAP-based Toffoli decomposition.---}
So far, we have described Toffoli decompositions suited for fixed-frequency qubits that use cross-resonance CR$_\theta$ for a two-qubit gate with CR$_{\pi/2}$ locally equivalent to CNOT.
In tunable-frequency qubits, a different class of two-qubit gates is possible, and the most popular gates used are the iSWAP ($=\ket{0}\bra{0} + i \ket{0}\bra{1} + i \ket{1}\bra{0} + \ket{1}\bra{1}$) gate and the controlled-Z gate (CZ), which is locally equivalent to CNOT.
Variants of the Toffoli gate decomposed into these two-qubit gates were first demonstrated by using an elaborate flux-pulse sequence in Refs.~\cite{fedorov2012,reed2012realization}.
Here we show in Fig.~\ref{fig3}(b) a more simple iSWAP-based decomposition derived from a similar spirit to the ones in Fig.~\ref{fig2} that uses only two $X_{12}$ gates as qutrit ones, which deactivate the iSWAPs when the first control qubit Q1 is $\ket{1}$.
The iSWAPs prevent the two control inputs of $\ket{01}$ from working on the CX, giving a total phase of $\pi$ that can be compensated by adding a two-qubit gate, for example, a $\ket{0}$-controlled Z gate locally equivalent to CZ.

\paragraph{Application to real-time QEC based on imcomplete Toffoli gates.---}
In QEC, how to reset ancilla qubits is an important task, and there have been developed several ways to reset transmon states actively beyond passive waiting~\cite{riste2013,salathe2018,andersen2019,bultink2020,riste2020,corcoles2021}.
In the conventional QEC, ancilla qubits are read out with projective measurements, and if the state is projected to $\ket{1}$, conditional electronics issues an $X$ gate.
This procedure is repeated a few times to confirm until the state is projected to $\ket{0}$.
On the other hand, to reinitialize the ancillae, here we use an all-microwave approach by applying the resonator-assisted double drive reset (DDR) scheme~\cite{magnard2018,geerlings2013}~(Fig.~\ref{fig3}(c)).
Two simultaneous drives $D1$ and $D2$ on a transmon couple the qutrit excited states $\ket{{\rm Q}=1,n=0}$ and $\ket{{\rm Q}=2,n=0}$ to $\ket{{\rm Q}=0,n=1}$, whose resonator photon decays through a lossy channel with the rate $\kappa$.
Apart from the original design of resonator-assisted DDR, the capability to reset $\ket{1}$ and $\ket{2}$ can be utilized in our QEC.

\begin{figure}[t]
\includegraphics[width=87mm]{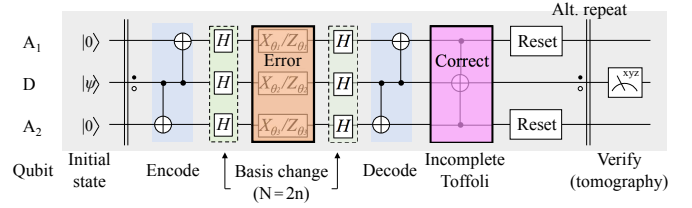}
\caption{
Measurement-free real-time error correction.
The initial state of a data qubit (D) is encoded into two ancillae (A1 and A2) with CNOT gates.
The bit-flip errors on the three qubits are decoded and corrected by the incomplete Toffoli gate that leaves the ancillae in qutrit states, whose excited states ($\ket{1}$ and $\ket{2}$) are reinitialized to $\ket{0}$ by the resonator-assisted double drive reset.
The protocol is alternatingly repeated with basis change and revert for the correction of phase-flip errors.
\label{fig4}}
\end{figure}

Figure~\ref{fig4} shows the three-qubit QEC protocol.
The state of a data qubit ($\ket{\psi} = \alpha \ket{0} + \beta \ket{1}$) is encoded into two ancilla qubits as $\alpha \ket{000} + \beta \ket{111}$ with $\ket{{\rm Q_1 Q_2 Q_3}}=\ket{{\rm A_1 D A_2}}$.
A bit-flip error on a qubit is expressed by the rotation about the $X$-axis as $X_{\theta_{i}}$ with $\theta_{i}$ ($i=1,2,3$) the error rotation angle.
An error on the data qubit ($X_{\theta_2}$) changes the three-qubit state into
\begin{equation}
\begin{split}
\alpha \left(\cos({\theta_2/2}) \ket{000} - i\sin({\theta_2/2})\ket{010} \right) \\+ \beta \left(\cos({\theta_2/2})\ket{111} - i\sin({\theta_2/2})\ket{101} \right).
\end{split}
\end{equation}
The errors are then decoded to
\begin{equation}
\begin{split}
\cos({\theta_2/2}) (\alpha \ket{000} + \beta \ket{010}) \\-i\sin({\theta_2/2}) (\alpha \ket{111} + \beta \ket{101})
\end{split}
\end{equation}
and corrected by the Toffoli gate to
\begin{equation}
\ket{\psi} \otimes \left(\cos({\theta_2/2})\ket{00} -i\sin({\theta_2/2})\ket{11} \right),
\end{equation}
with the ancilla state $\ket{{\rm A_1 A_2}}$ separated.
Similarly, an error on A1 or A2 produces a three-qubit state after the correction of $\ket{\psi} \otimes \left(\cos({\theta_1/2})\ket{00} -i\sin({\theta_1/2})\ket{01} \right)$ or $\ket{\psi} \otimes \left(\cos({\theta_3/2})\ket{00} -i\sin({\theta_3/2})\ket{10} \right)$, respectively.

In our Toffoli decompositions, the gate sequence is inverted at the central two-qutrit gate to conserve the input control states for a complete Toffoli functionality.
As the target operation is already accomplished at the central two-qutrit gate, stopping the gate sequence at this point leaves the control state in qutrit space.
Even if this incomplete variant of the Toffoli gate is used in our QEC instead of a complete one, the data qubit state $\ket{\psi}$ becomes separable, and the DDR can reinitialize the qutrit states of the ancillae as $\ket{\psi} \otimes \sum_{i,j=0}^2 c_{ij} \ket{ij} \rightarrow \ket{\psi} \otimes \ket{00}$ up to a global phase.
Thus, it significantly reduces the length of the Toffoli gate sequence; especially the number of CX gates required decreases to two.

The capability to correct phase-flip errors is obtained by changing and reverting the basis with Hadamard gates.
The protocol can correct both bit- and phase-flip errors by alternatingly repeating the error correction with and without the basis change.

\paragraph{Discussion.---}

\begin{table}[!t]
\caption{
Breakdown of the time required for the error correction and ancilla reset in measurement-based (MB) and measurement-free (MF) approaches.
The incomplete Toffoli gate B3 in Fig.~\ref{fig2} is assumed for the MF correction gate.
Note that there is a sizeable overlap between the kernel integration and the branch determination times.
}
\label{tab1}
\begin{ruledtabular}
\small
\begin{tabular}{lcc}
& MB (Ref.) &  MF (this work)     \\ \hline
Latency (cable + electronics) & 160 ns & NR \\
Kernel integration & 320 ns & NR \\
Resonator emptying & 260 ns & NR \\
Branch determination & 420 ns & NR \\
Single-qubit/qutrit gate & 30 ns & 30 ns \\
Cycle multiplicity & 2 & NR \\
Two-qutrit gate (Q1-Q2) & NR & 90 ns~\cite{wei2021} \\
Exclusive CNOT gate (Q2-Q3) & NR & 125 ns~\cite{blok2021} \\
Double drive qutrit reset & NR & 280 ns ($<1$\%)~\cite{magnard2018} \\ \hline
Total & 1.4~${\rm \mu}$s & 525 ns \\
\end{tabular}
\leftline{NR: not relevant, Ref.: Reference~\cite{corcoles2021}}
\end{ruledtabular}
\end{table}

The time required for the error correction and ancilla reset in measurement-based and measurement-free approaches is shown in Tab.~\ref{tab1}.
From the availability of those typical values, we refer to Ref.~\cite{corcoles2021} as the measurement-based approach that uses projective measurements on the ancilla qubits and conditional reset and correct pulses to ancilla and data qubits, respectively.
In the measurement-based approach, there is an inevitable delay of 100-200~ns between the issue of a readout pulse at room temperature electronics and the arrival of the signal returned from the device.
The signal is then integrated for $\sim300$~ns at the room temperature electronics to determine the qubit state and decide whether to issue the reset and correction pulses to the qubits (branch determination).
During this process, the resonators are emptied for $\sim300$~ns after their population.
This cycle is repeated a few times to obtain a higher reset fidelity.

For the measurement-free error correction gate, we assume the incomplete Toffoli gate B3 in Fig.~\ref{fig2}, composed of a controlled qutrit gate on Q1-Q2, a single qutrit gate $X_{12}$ on Q2, and a pure (exclusive) CNOT gate on Q2-Q3 that keeps the target Q3 state for ${\rm Q2}=\ket{2}$.
As the first Q1-Q2 gate does not require such exclusivity, a fast and high fidelity cross-resonance gate of about 90~ns can be obtained by continuously driving the two capacitively-coupled qutrits to suppress their ZZ coupling~\cite{wei2021,mitchell2021}.
The deterministic qutrit reset with the resonator-assisted DDR scheme has demonstrated a reset time of 280~ns with infidelity less than 1\%~\cite{magnard2018}.
Furthermore, a shorter reset time of $\sim 80$~ns would be possible by optimizing the resonator's photon emission rate $\kappa$, transmon's anharmonicity $\alpha$, and the transmon-resonator coupling $g$~\cite{walter2017,magnard2018}.

In comparison, about a three-fold reduction of a total gate time for the correction and reset of transmon states can be envisioned (Tab.~\ref{tab1}), enabling a fast repetition rate of $\gtrsim 1$~MHz.
Notably, the total time in the measurement-based approach is limited by multiple sources of similar size: latency, signal integration (statistics), readout resonator's response, and the process and determination speed of classical electronics (FPGA).
On the other hand, the major limiting factors of our all-microwave scheme are different from them: gate times and reset resonator's coupling to the environment.
Eventually, transmons dedicated for the QEC ancillae do not require readout resonators, lines, and electronics.
Thus, they can significantly reduce the hardware resource necessary for QEC, which is another benefit of this scheme.

\paragraph{Conclusion.---}
We have described a simple and efficient way of realizing multi-controlled gates by using higher-dimensional state space as an additional computational resource and derived a series of novel Toffoli decompositions in the first part.
Several important concepts were introduced to quantify their differences from the other Toffoli gate implementations;
(1)~scalability to a generalized Toffoli gate with an arbitrary number of control qubits,
(2)~the exclusivity on the control state in the central CNOT that works on the target qubit for a specified control state and does not for the other qutrit states,
(3)~incomplete variants of Toffoli decompositions that can drop the uncomputation part right after the target operation to reduce the number of two-qutrit gates to two for a particular usage, such as QEC, and
(4)~minimal $\ket{2}$-state decomposition that minimizes the time during which the controls stay in the $\ket{2}$ state among equivalent or hybrid decompositions.

As an important application of the Toffoli decompositions, we proposed a novel three-qubit QEC protocol that works without projective measurements and classical feedforward.
Combined with an all-microwave reset scheme and fast and high-fidelity two-qubit gates, it can provide real-time QEC with a repetition rate of $\gtrsim1$~MHz for bit and phase flip errors.
Since the series of Toffoli gate decompositions shown here can be linearly extended to have an arbitrary number of control qubits, it may pave the way for implementing more practical QEC protocols in the near term.

\paragraph{Acknowledgments.---}
We acknowledge
Naoki Kanazawa,
Masao Tokunari,
Koji Masuda,
Daiju Nakano,
Matthias Steffen,
Jerry M. Chow,
and Kouichi Semba,
for helpful discussions.
This work was supported in part by Japan Society for the Promotion of Science (JSPS), through Grants-in-Aid for Scientific Research (KAKENHI) Grant No.~JP20H01911, JP20K22347, and 21K19761, and by
U.S.-Japan Science and Technology Cooperation Program in High Energy Physics.

\bibliography{cite.bib}

\clearpage

\onecolumngrid
\begin{center}
    \textbf{\large Supplementary Material for\\ 
$\;$\\ $\;$
``Measurement-Free Ultrafast Quantum Error Correction by Using Multi-Controlled Gates in Higher-Dimensional State Space"}
\end{center}
\bigskip

\input{sup}

\end{document}

%% file: sup.tex





\begin{figure}[b!]
\includegraphics[width=120mm]{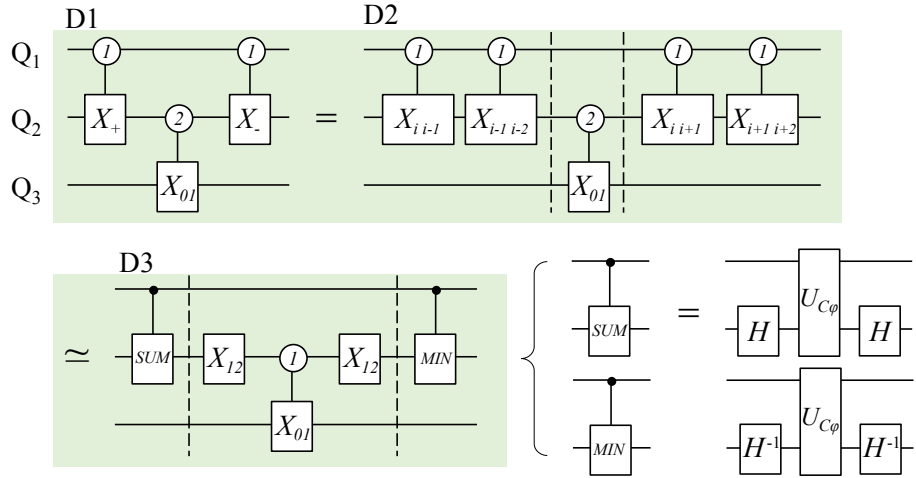}
\caption{
D1:~Original Toffoli gate decomposition with two controlled rotational gates (X$_+$ and X$_-$)~\cite{gokhale2019}.
Similar decompositions are possible, for example, by swapping the two rotational gates, yielding a NOT operation on the target when the input state is $\ket{{\rm Q_1 Q_2}}=\ket{10}$.
D2:~Each controlled rotational gate in D1 is naively decomposed into two controlled subspace gates, producing five two-qutrit gates in total.
Note that ``mod~3'' is abbreviated in the subscripts of the Q$_1$-Q$_2$ gates for simplicity.
D3:~More elaborate decomposition with three two-qutrit gates by using non-trivial single- and two-qutrit gates (see the main text for details).
}
\label{fig5}
\end{figure}

Qutrit-based Toffoli gate decomposition was first proposed in Ref.~\cite{gokhale2019} by using a $\ket{2}$-controlled $X_{01}$ gate sandwiched by a pair of $\ket{1}$-controlled $X_{+}$ and $X_{-}$ gates, as shown in D1 of Fig.~\ref{fig5}.
Note that similar decompositions are possible, for example, by swapping the two rotational gates, yielding a NOT operation on the target when the input state is $\ket{{\rm Q_1 Q_2}}=\ket{10}$.
Since the rotational gates are decomposed into subspace gates as
\begin{eqnarray}
X_{+} &=& X_{i\,j} X_{j\,k}\ \ \ \ {\rm with}\ \,j=i-1\ {\rm mod}\ 3,\ \,k=i-2\ {\rm mod}\ 3   \\
X_{-} &=& X_{i\,j'} X_{j'\,k'}\ \ {\rm with}\ j'=i+1\ {\rm mod}\ 3,\ k'=i+2\ {\rm mod}\ 3,
\end{eqnarray}
for $i=0,1,2$ (and with $X_{il} = X_{li}$), D1 can be transformed into D2 in Fig.~\ref{fig5}, where the modulus is abbreviated for simplicity.
As D2 contains five two-qutrit gates, this naive decomposition costs unreasonably high in the actual implementation.

A more elaborate implementation of D1 is possible by directly approximating the rotational gates to a controlled SUM or controlled MINUS gate~\cite{blok2021}, as shown in D3 of Fig.~\ref{fig5}.
They work on $\ket{{\rm Q_1Q_2}}$ as
\begin{eqnarray}
{\rm CSUM}\ket{m, n} &=& \ket{m, n+m \ {\rm mod} \ 3} \\
{\rm CMIN}\ket{m, n} &=& \ket{m, n-m \ {\rm mod} \ 3},
\end{eqnarray}
which are composed of the controlled phase gate $U_{{\rm C}\phi} = \sum_{n=0}^{2} \ket{n} \bra{n} \otimes Z^{n}$ with qutrit $Z = \sum_{j=0}^{2} e^{i 2\pi/3 j} \ket{j}\bra{j}$,
sandwiched by a pair of qutrit Hadamard gates
\begin{equation}
H = \frac{1}{\sqrt{3}}
\begin{pmatrix}
1&1&1\\
1&e^{i 2\pi/3}&e^{-i 2\pi/3}\\
1&e^{-i 2\pi/3}&e^{i 2\pi/3}\\
\end{pmatrix},
\end{equation}
implemented with a four-pulse sequence~\cite{morvan2021} or simultaneous driving of all three transitions~\cite{yurtalan2020}.
Although the decomposition D3 consists of three two-qutrit gates, the costs for those non-trivial gates are also high, supporting the efficiency of decompositions into controlled subspace gates described in the main text.

